# Motor-driven restructuring of cytoskeleton composites leads to tunable time-varying elasticity

Janet Y. Sheung[1]*, Daisy H. Achiriloaie[1], Christopher Currie[2], Karthik Peddireddy[2], Aaron Xie[1], Jessalyn Simon-Parker[1], Gloria Lee[2], Michael J. Rust[3], Moumita Das[4], Jennifer L. Ross[5], Rae M. Robertson-Anderson[2]*

[1] W. M. Keck Science Department, Scripps College, Pitzer College, and Claremont McKenna College, 925 N. Mills Ave. Claremont CA 91711, USA
[2] Department of Physics and Biophysics, University of San Diego, 5998 Alcala Park, San Diego, CA, 92110, USA
[3] Department of Molecular Genetics and Cell Biology, University of Chicago, Chicago, IL 60637, USA.
[4] School of Physics and Astronomy, Rochester Institute of Technology, Rochester, NY 14623, USA.
[5] Department of Physics, Syracuse University, Syracuse, NY 13244, USA.

* jsheung@kecksci.claremont.edu, randerson@sandiego.edu

## ABSTRACT

The composite cytoskeleton, comprising interacting networks of semiflexible actin and rigid microtubules, generates forces and restructures using motor proteins such as myosins to enable key processes including cell motility and mitosis. Yet, how motor-driven activity alters the mechanics of cytoskeleton composites remains an open challenge. Here, we perform optical tweezers microrheology and confocal imaging of composites with varying actin-tubulin molar percentages (25-75, 50-50, 75-25), driven by light-activated myosin II motors, to show that motor activity increases the elastic plateau modulus by over two orders of magnitude by active restructuring of both actin and microtubules that persists for hours after motor activation has ceased. Nonlinear microrheology measurements show that motor-driven restructuring increases the force response and stiffness and suppresses actin bending. The 50-50 composite exhibits the most dramatic mechanical response to motor activity, due to the synergistic effects of added stiffness from the microtubules and sufficient motor substrate for pronounced activity.

The cytoskeleton is an active composite of protein filaments capable of restructuring on demand to meet the wide variety of mechanical properties needed by eukaryotic cells, such as structural rigidity and malleability.[1–4] Semiflexible actin filaments, rigid microtubules, and intermediate filaments all contribute to this versatility by forming interacting viscoelastic networks.[2,5–8] In addition, molecular motors, such as actin-associated myosins, stochastically bind and actively pull on filaments, generating forces to contract and rearrange the cytoskeleton.[9–12] Motor-driven active dynamics play a critical role in enabling the cytoskeleton to rapidly tune its mechanical properties to achieve a myriad of different functions in response to environmental cues. At the same time, interactions between actin and microtubules play equally important roles in processes such as cell division, migration, and wound healing.[1,13–17] As such, the cytoskeleton is an exemplar of active matter that has been intensely studied not only for its biological relevance but for its importance in the design of active non-equilibrium materials.[18,19,12,20,21] Nevertheless, the mechanical properties of the active composite cytoskeleton, remain poorly understood.

Numerous in vitro studies have been carried out on actomyosin systems, reporting a range of structural and dynamical properties depending on the concentrations of actin, myosin and crosslinkers.[22,10,23,24,21,25] With sufficiently high crosslinker concentration, actomyosin activity induces large-scale contraction and coarsening of disordered actin networks.[26,22,27–29,24] However, at lower crosslinker densities, networks undergo destabilizing flow and rupturing into disconnected foci, thereby weakening the network.[22,27] We previously showed that incorporating microtubules into an actomyosin network provides a mechanical scaffold that enables controlled contractile dynamics, without flow or rupturing, in the absence of crosslinkers.[9] We also showed, surprisingly, that both actin and microtubules exhibited ballistic motion, with speeds that were indistinguishable from one another.[9,30] This result differs from that of steady-state composites of crosslinked actin and microtubules in which the mobilities of actin and microtubules were distinct.[22]

The rheology of actomyosin systems is far less understood, primarily due to the fact that microrheology methods typically used to investigate similar biological systems rely on the generalized Stokes Einstein relation (GSER) to derive viscoelastic moduli from thermal fluctuations of embedded particles.[21] In non-equilibrium active systems, GSER is violated over frequency ranges comparable to those of motor-driven activity.[22] Previous attempts to investigate the mechanics of actomyosin systems have shown that particle fluctuations are dominated by motor activity at low frequencies while thermal fluctuations dominate the high-frequency regime.[23] However, how interactions between actin and microtubules impact the rheology of actomyosin systems remains completely unexplored.

We previously showed that the mechanics of steady-state actin-microtubule composites can be tuned by varying the relative concentrations of actin, microtubules and crosslinkers.[22,27,31] Notably, we found that increasing microtubule concentration suppressed actin bending and led to more elastic-like response, while increasing actin concentration decreased mechanical heterogeneity by reinforcing microtubules against buckling.

Here, we perform linear and nonlinear optical tweezers microrheology to characterize the microscale force response of myosin-driven actin-microtubule composites before, during and following myosin activity (Fig 1, Section S1). We circumvent GSER violation by performing measurements on judiciously tuned composites, with actin-tubulin molar percentages of 25-75, 50-50 and 75-25, which visibly restructure on the order of minutes, yet are still slow enough ($\omega_{active}$<0.2 rad/s) to be considered quasi-static (i.e., dominated by thermal fluctuations) for frequencies above ~rad/s. Further, we use light-sensitive myosin inhibitor, blebbistatin, to precisely control the location and duration of motor activity, enabling measurements before, during and at precise time points after motor activity. We complement our measurements with confocal fluorescence imaging to show that the increased elasticity we measure, tuned by the molar percentages of actin and tubulin, results from mesoscale restructuring of homogeneous entangled meshes to more sparsely connected clusters and bundles.

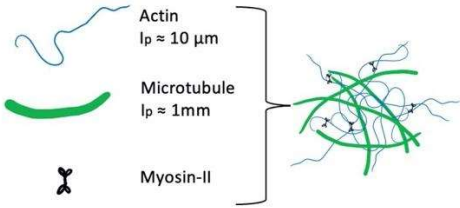
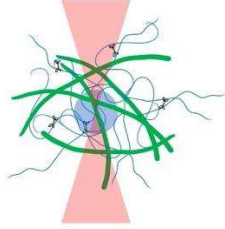
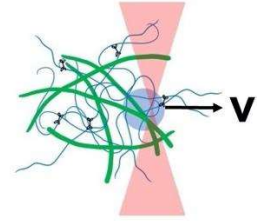
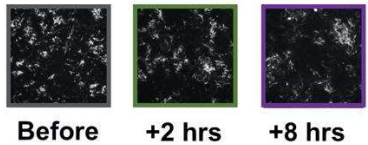
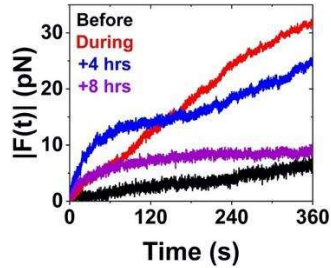
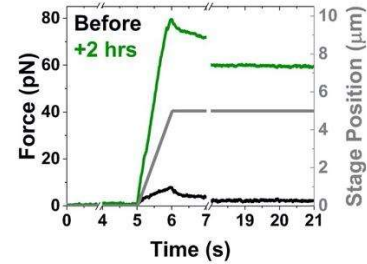

**Figure 1. Optical tweezers microrheology characterizes how actomyosin activity alters the mechanical response of actin-microtubule composites.** (A) Cartoon of composite comprised of actin filaments, microtubules, and myosin II minifilaments, prepared as described in SI. (B) 512x512 pixel (213µm x 213µm) fluorescence confocal images of rhodamine-labeled microtubules in a 50-50 actin-tubulin composite acquired before, 2 hrs after, and 8 hrs after myosin activation. (C) Linear Microrheology. (Top) Cartoon depicting trapping a 4.5-$\mu$m diameter microsphere within a composite and tracking its force fluctuations $(F_x, F_y)$ for 3 mins. (Bottom) Sample force magnitudes $|F(t)| = (F_x^2 + F_y^2)^{1/2}$ measured for the 50-50 composite before (black), during (red), and 4 hrs (blue) and 8 hrs (purple) after myosin activation. (D) Nonlinear Microrheology. (Top) Cartoon of measurement in which we hold an optically trapped bead fixed for 5 s, displace the bead 5 $\mu$m through the composite at 5 $\mu$m/s, then hold it fixed again for the remaining of the 20 s measurement. We measure the force exerted on the microsphere for the duration of the measurement. Example force measurement for the 50-50 composite before (black) and 2 hrs after (green) 10 mins of myosin activation. The grey curve shows the position of the piezoelectric stage that displaces the bead relative to the composite.

To characterize the linear force response, we measure the force fluctuations $(F_x, F_y)$ of optically trapped beads within the composites before, during, 4 hrs after, and 8 hrs after myosin activation (via blebbistatin deactivation) (Fig 1C, Fig S2). The magnitude of the fluctuations $|F(t)| = (F_x^2 + F_y^2)^{1/2}$ increases during and following myosin activity for all composites, with 50-50 exhibiting the most pronounced increase. Interestingly, while we see no visible signs of contraction, on the order of minutes, after we halt activation, all composites continue to change their mechanical properties for at least 8 hrs following activation. In fact, in all cases, the maximum $|F(t)|$ was largest 4 hours after motor activation ceased, rather than during active contraction (Fig

S2). At this time, the force exerted on the trapped bead by the 50-50 composite increased beyond the strength of the trap in some cases as indicated by the arrows in Fig 2A.

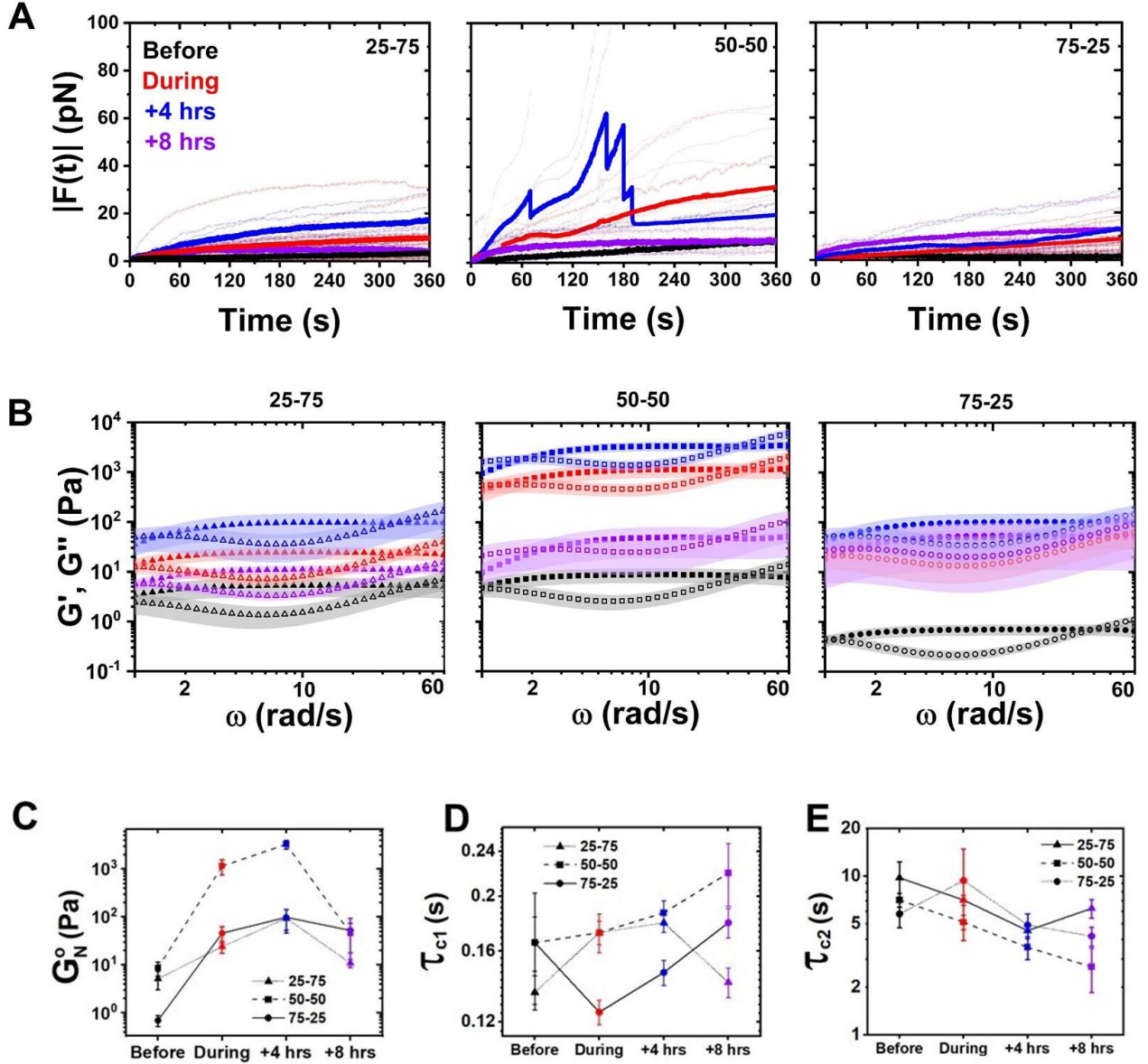

**Figure 2. Myosin activity universally increases the elastic plateau of actin-microtubule composites with varying compositions.** (A) Magnitude of the force exerted on optically trapped microspheres, $|F(t)| = (F_x^2 + F_y^2)^{1/2}$, embedded in composites of varying actin-tubulin molar percentages (listed in each panel) measured before (black) and during (red) 3 minutes of myosin activation, as well as 4 hours (blue) and 8 hours (purple) after. Lighter dotted lines indicate individual trials and darker solid lines indicate averages of the individual trials shown. The force exerted by the 50-50 composite at +4 hrs exceeded the strength of the optical trap for some trials, leading to the truncated force curves shown in the 50-50 plot. Arrows above the average force curve during activity indicate locations where the number of individual trials contributing to the average decreases due to truncated trials. (B) Elastic ($G'(\omega)$, closed symbols) and viscous ($G''(\omega)$,

open symbols) moduli computed from the forces shown in A. Color coding and panel organization is as in A. The shaded region surrounding each curve indicates standard error. (C) Elastic plateau modulus $G_N^0$ determined from the data shown in B. (D,E) Fast ($\tau_{c1}$, D) and slow ($\tau_{c2}$, E) relaxation times determined from the high ($\omega_{c1}$) and low ($\omega_{c2}$) frequencies at which $G'(\omega)$ and $G''(\omega)$ cross.

From the force fluctuations we determine the frequency-dependent storage and loss moduli, $G'(\omega)$ and $G''(\omega)$, using GSER as previously described (SI Methods, Fig 2C).[32–34] Because GSER is only strictly valid for steady-state thermal systems, we restrict our analyses to frequencies well above the active driving frequency $\omega_{active}$, which we determine using our previously measured contraction velocity $v \approx 35$ nm/s for the fastest composite (75-25).[30] We compute $\omega_{active} = 2\pi\dot{\gamma}_{active} \approx 0.18$ rad/s, and thus safely approximate a quasi-steady-state for $\omega > 1$ rad/s ($>5\omega_{active}$).[35]

As shown in Fig 2B, the viscoelastic moduli for all composites and time points exhibit strong entanglement behavior with the storage modulus $G'(\omega)$, a measure of elastic storage, exhibiting a plateau $G_N^0$, and $G''(\omega)$, the dissipative component, exhibiting a local minimum.[36] With motor activity, the magnitude of the elastic plateau $G_N^0$ increases by 1-2 orders of magnitude for all composites and continues to increase up to 4 hrs after activation, after which it drops but maintains a higher value compared to pre-activation. This drop at 8 hrs is most dramatic for 50-50 while nearly undetectable for 75-25. As $G_N^0$ is predicted to scale linearly with the number of entanglements along each polymer, this non-monotonic dependence of $G_N^0$ on time, relative to motor activation, signifies a substantial increase in entanglement density followed by relaxation back to a less densely entangled mesh.[36]

Nearly all composites exhibit crossovers of $G'(\omega)$ and $G''(\omega)$ at both low and high frequencies (Fig 2B). For entangled polymers, the high crossover frequency, $\omega_{c1}$, is a measure of the entanglement time $\tau_e$, i.e., the time at which filament-filament interactions become important;[28] while the low crossover frequency, $\omega_{c2}$, is a measure of the tube model disengagement time $\tau_D$, i.e., the time at which filaments curvilinearly diffuse out of their confinement tubes. We measure $\tau_{c1} = \frac{2\pi}{\omega_{c1}}$ values of ~0.14-0.24 s, comparable to the predicted entanglement time of ~0.25 s for an actin network at the same concentration.[37,38] As previously shown and predicted,[37,20] the lower values we measure after activity are likely due to actin bundling which decreases $\tau_{c1}$ as bundles become stiffer and more connected.

We measure $\tau_{c2} = \frac{2\pi}{\omega_{c2}}$ values of ~4–9 s for all composites, quite similar to the previously reported longest relaxation timescales of ~3-9 s for comparable actin-microtubule composites that lacked motors.[27] As in this previous work, we attribute the relatively fast disengagement time, compared to predictions,[37,38] as arising from entanglement hopping, whereby filaments can periodically move transversely to their contours due to temporary release of an entanglement with a neighboring filament.[31,38,39] We note that $\tau_{c2}$ generally decreases (albeit with substantial noise) for all composites during and following activation, which may arise from more hopping events as the network restructures, releasing and reforming entanglements.

To determine the extent to which motor-driven viscoelastic changes are preserved when the composite is driven far from steady-state, we perform nonlinear constant-rate strain measurements before and 2 hrs after motor activation. Specifically, we measure the force exerted on an optically trapped bead as we displace it at a constant speed and subsequently hold it fixed to measure the force relaxation (Fig 1D). We also attempted this protocol at 4 hrs after activation but the 50-50 composite was too stiff to complete most of the measurements without the bead being pulled out of the trap, consistent with the continued increase in $G_N^0$ (Fig 2).

As shown in Fig 3A, the force response both before and after activity is solid-like (i.e., linear strain dependence) over the entire strain for all composites, but the force magnitudes and slopes all increase after activity. While the maximum force increases by >2x for 50-50 and 75-25, with 50-50 exhibiting the highest resistive force, it only modestly increases for 25-75. In all composites, the increase is a lower bound that does not include the trials in which the trapping force could not withstand the composite force (indicated by single data points in Fig 3A). To quantify the increased slope, a measure of composite stiffness, we compute the strain-averaged differential modulus $K = dF/dx$ (Fig 3A inset). As shown, $K$ values follow a similar trend as the force magnitude, suggestive of more bundled filaments that are stiffer and more readily resist strain.

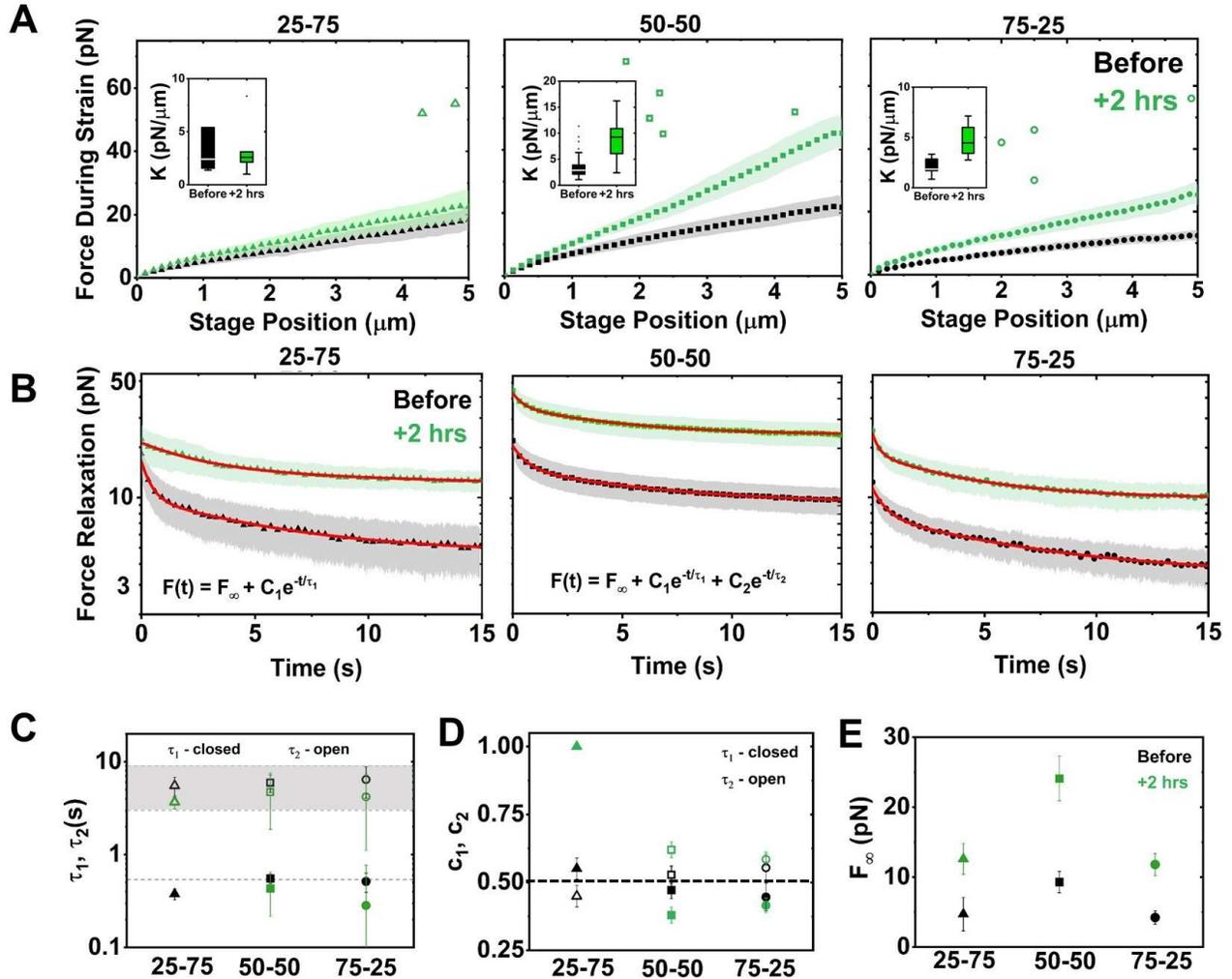

**Figure 3. Myosin-driven restructuring increases the nonlinear force response and suppresses relaxation of actin-microtubule composites**. (A) Average force $F(x)$ exerted on the bead during nonlinear strain, measured before (black) and 2 hrs after (green) myosin activation, for composites with actin-tubulin percentages of 25-75 (left), 50-50 (middle) and 75-25 (right). Shaded regions along each curve indicate standard error. Insets: Boxplots of differential modulus, $K = dF/dx$, found as the slope of $F(x)$ for each trial. (B) Relaxation of force versus time following strain, measured before and 2 hours after motor activation. Nearly all curves are well-fit to a sum of two exponentials and a non-zero offset: $F(t) = F_\infty + C_1 e^{-t/\tau_1} + C_2 e^{-t/\tau_2}$. The 25-75 relaxation after activation fits to a single exponential. Fits are shown as red lines. (C) Time constants corresponding to fast ($\tau_1$, closed symbols) and slow ($\tau_2$, open symbols) modes determined from fits in B for relaxations measured before (black) and 2 hrs after (green) activation. Dashed line corresponds to predicted actin bending timescale $\tau_B$ and gray region corresponds to previously measured reptation times for steady-state composites. (D) Relative contributions of the fast ($c_1$) and slow ($c_2$) modes determined from the fits in B. Dashed horizontal line indicates equal contribution from both

modes. (E) $F_\infty$ determined from the fits shown in B, corresponding to the force that is sustained at the end of the relaxation period.

The force relaxation curves following strain (Fig 3B,C) show that both before and 2 hrs after activation, all composites maintain some degree of strain memory (i.e., non-zero force) at the end of the relaxation phase. However, this non-zero terminal force $F_\infty$, a signature of solid-like mechanics, is >2x higher following myosin activation for all composites, in line with our previous measurements on statically crosslinked actin-microtubule composites (Fig 3B).[22,27] Notably, while the force during strain for 25-75 does not increase significantly after motor activity, the stress relaxation is markedly suppressed. Considering $F_\infty$, we find that nearly all relaxations fit well to a sum of two exponentials with well-separated time constants: $F(t) = F_\infty + C_1 e^{-t/\tau_1} + C_2 e^{-t/\tau_2}$. We interpret each decay as arising from a distinct relaxation mechanism with a characteristic decay time $\tau_i$ and a relative contribution to the stress relaxation $c_i = C_i/(C_1 + C_2)$ (Fig 3C,D). The exception is 25-75 after activity, which is accurately described by a single exponential that lacks the fast timescale $\tau_1$.

Our slow relaxation timescales are similar to our measured $\tau_{C2}$ values before and after activity for all composites (Fig 2E), and likewise modestly decrease following motor activity. As such we understand this timescale as arising from reptation and hopping, facilitated by restructuring.[31] To understand our fast relaxation timescale, $\tau_1$, we evaluate the predicted actin bending timescale, $\tau_B \approx \gamma L/(l_p k_b T (3\pi/2)^4)$ where $L \approx 8.7 \pm 2.8$ μm is the average actin filament length,[31,38,39] $l_p \approx 17$ μm the persistence length,[40] and $\gamma = 4\pi\eta/ln(2\xi/r)$ the drag coefficient, with $\eta$ the solvent viscosity, $\xi \approx 0.85$ μm the actin mesh size,[41] and $r \approx 3$ nm the filament radius[42]. Our computed value $\tau_B \approx 0.54 \pm 0.17$ s (shown as dashed line in Fig 3C) is quite close to our measured fast timescales (Fig 3C).

By evaluating the relative contribution of each relaxation mechanism ($c_1$, $c_2$) before and after activity (Fig 3C), we find that, in all cases, the slow mode contributes more to the relaxation than the fast mode ($c_2 > c_1$). Following motor activation, the fast bending contribution decreases further, or is eliminated, and $c_2$ concomitantly increases. Motor activity may suppress bending by bundling filaments, reducing their susceptibility to thermal bending.[43] The lack of $\tau_1$ in 25-75 suggests that actin bundling, along with the larger percentage of microtubules is sufficient to suppress bending entirely. Notably, these results are distinct from our results for statically crosslinked composites, in which the fast mode contributed >2x more than the slow mode, even with substantial crosslinking.[22]

We hypothesize that the rheological behavior shown in Figs 2 and 3 is due to restructuring of composites from meshes of individual filaments to networks of bundles and clusters.[37,44,45] To test this hypothesis, we analyze confocal fluorescence images of rhodamine-labeled microtubules in composites at the same time points that we perform microrheology (Fig 4, Fig S1).

The images show clear restructuring of all composites over 8 hrs following myosin activation (Fig 4A), which we quantify by computing the spatial image autocorrelation function $g(r)$ (see SI Methods), which determines the degree to which the intensity at one location in an image correlates with the intensity of the surrounding points at varying distances.[46] The slower $g(r)$ decays with increasing $r$, the larger the structural features of the network. Average $g(r)$ curves exhibit clear broadening over time after motor activation, as shown for 50-50 in Fig 4B, suggesting formation of clusters separated by larger, sparsely populated gaps. By fitting $g(r)$ to an exponential $g(r) = Ae^{-\frac{r}{\xi}}$, we determine a characteristic decay length or correlation length $\xi$, which scales as the mesh size for an isotropic network.[47] As shown in Fig 4C, the magnitude and spread of $\xi$ values increases with time following activation for all composites, consistent with the restructuring we describe above.

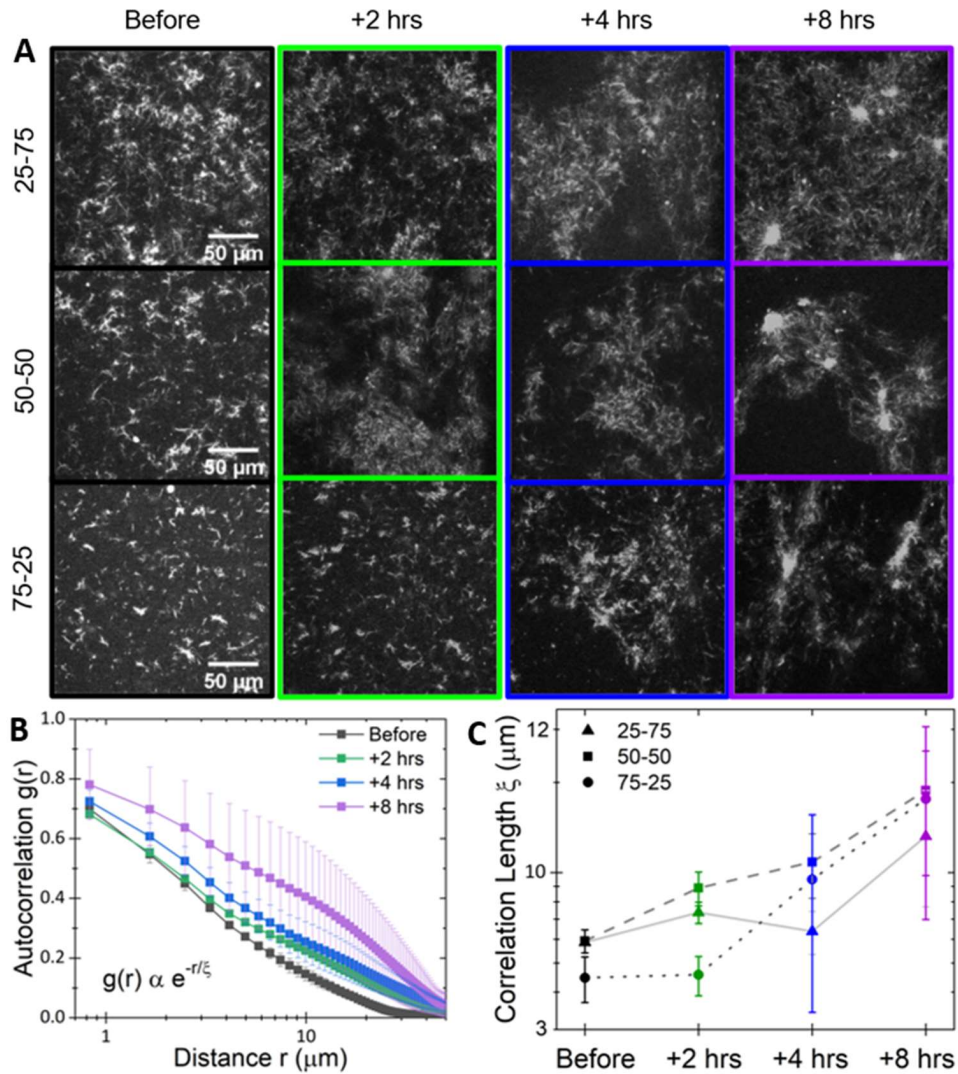

**Figure 4. Myosin activity drives sustained mesoscale clustering in actin-microtubule composites.**
(A) Representative 512x512 pixel fluorescence confocal micrographs of rhodamine-labeled microtubules in actin-microtubule composites with actin-tubulin molar percentages of 25-75 (top), 50-50 (middle), and 75-25 (right). Images were acquired with 568 nm illumination at varying times relative to myosin activation: Before (black), +2 hrs (green), +4 hrs (blue) and +8 hrs (purple) after. (B) Average autocorrelation curves $g(r)$, computed from five images for each time point, for the 50-50 composite. Error bars are standard error. Curves are fit to the equation shown to determine the correlation length $\xi$ for each composite and time shown in C. (C) Average correlation lengths $\xi$ for 25-75 (triangles), 50-50 (squares) and 75-25 (circles) composites at each time shown in A.

Importantly, the mechanical properties we report here are distinct from those expected for actomyosin networks without microtubules. Namely, in actomyosin networks without crosslinkers, motor activity causes network rupturing and fluidization. Such activity would result in reduced elasticity as well as unstable structure following activity. Here, microtubules act as a scaffold that reinforces and further connects the actin, allowing myosin to contract the network into clusters while maintaining connectivity and mechanical integrity over extended periods and when subject to nonlinear forcing.[9] Further, we show that the 50-50 composite surprisingly exhibits the highest strength and elasticity (rather than 25-75 that has more rigid microtubules) as well as the strongest effect of motor activity (rather than 75-25 that has more actin for myosin to act on). While composites with more actin (75-25) have more active substrate, they are inherently floppier and require more contraction to achieve the same stiffness and rigidity that 50-50 confers. Likewise, 25-75 is stiffer from the outset, but has less active substrate, so is less susceptible to motor-driven restructuring.

In summary, we couple optical tweezers microrheology with fluorescence confocal microscopy to elucidate the effects of myosin II activity on the mechanics and structure of entangled actin-microtubule composites. We show that motor activity increases the plateau modulus, resilience to nonlinear straining, and mechano-memory; and suppresses filament bending during motor activity and for hours after cessation of motor activation. This time-varying mechanical response is due to slow restructuring from an entangled mesh of individual filaments to a network of dense bundles. Our results provide valuable insight into the mechanics of active cytoskeletal systems, and how the interplay between motor activity, the composite nature of the cytoskeleton, and time-varying structure leads to the myriad mechanical properties that cells exhibit. More generally, our techniques and results can be applied to a wide range of active matter systems currently under intense investigation.

**Experimental Section**

Many of the materials and methods are described in the main text and in the captions of Figs 1-4. More detailed descriptions of all materials and methods are included in the Supporting Information.

**Supporting Information**

Section S1. Materials and Methods

Figure S1. Confocal fluorescence images of microtubules in myosin-driven actin-microtubule composites show clustering and increased heterogeneity for hours after activation.

Figure S2. Data from Figure 2A plotted with log scaling on the y-axis.

**AUTHOR INFORMATION**

**Corresponding Author:** Rae M Robertson-Anderson

**Author Contributions.** R.M.R.-A. conceived the project, guided the experiments, interpreted the data, and wrote the manuscript. J.Y.S designed, performed, and helped guide the experiments, analyzed and interpreted the data and wrote the manuscript. D.H.A. helped perform experiments, analyzed data, prepared figures, and helped write the manuscript. C.C analyzed data, prepared figures, and helped write the manuscript. A.X. helped analyze data and prepared figures. J.S.-P. prepared figures. K.P. helped guide experiments and analyzed and interpreted data. G.L. helped with materials preparation and data interpretation. M.J.R., M.D. and J.L.R. helped conceive the project and interpret data.

**Acknowledgments.** This research was funded by a William M. Keck Foundation Research Grant and a National Institute of General Medical Sciences Award (no. R15GM123420). J. Y. S. acknowledges startup support from the W. M. Keck Science Department of Claremont McKenna, Scripps, and Pitzer Colleges of The Claremont Colleges. The authors are grateful to Jonathan Garamella and Ryan McGorty for helpful discussions; and Gregor Leech and Ryan Clairmont for assistance in instrument calibrations.